\documentclass[fleqn,twoside]{article}
\usepackage{epsfig}
\usepackage{amssymb}
\usepackage{rotating}
\usepackage{graphics}
\usepackage{graphicx}
\usepackage{espcrc2}

\def\a{\alpha}

\def\m{\mu}

\def\p{\pi}

\def\t{\tau}

\def\D{\Delta}
\def\F{\Phi}
\def\G{\Gamma}

\def\ve{\varepsilon}

\def\be{\begin{equation}}
\def\ee{\end{equation}}

\def\bea{\begin{eqnarray}}
\def\eea{\end{eqnarray}}

\def\pl#1#2#3{Phys.~Lett.~{\bf B {#1}} ({#2}) #3}

\title{
Flavor effects in thermal leptogenesis}

\author{Steve Blanchet\address[MPI]{Max-Planck-Institut f\"{u}r Physik,
 Werner-Heisenberg-Institut, F\"{o}hringer Ring 6, 80805 M\"{u}nchen}
and Pasquale Di Bari\addressmark[MPI]
}

\begin{document}

\begin{abstract}
We review recent developments in leptogenesis on flavor effects.
Their account discloses an important connection between the
matter-antimatter asymmetry of the Universe and $C\!P$ violation at low energies.
Besides, they modify the upper bound on the neutrino masses
holding in the unflavored case. In this respect, it is important to identify the exact
condition for flavor effects to be relevant and for the
`fully flavored' Boltzmann equations to be valid.
\end{abstract}

\maketitle

\section{Introduction}

In the context of leptogenesis~\cite{fy}, it has been recently
realized that the flavor composition of the leptons produced in
the decays of the heavy right-handed (RH) neutrinos, plays an
important role ~\cite{nardi,davidson1} (see also \cite{bcst,endoh,pilaftsis,vives}).

 The most striking consequence is the possibility that
 leptogenesis stems uniquely from the Majorana and Dirac phases
~\cite{nardi,davidson2,us2,many}.
This can happen because in flavored leptogenesis,
contrarily to the unflavored case,
the final baryon asymmetry depends also on the PMNS matrix and
on its phases. It turns out that even imposing {\it CP} invariance at
high energy, it is still possible to explain the observed baryon asymmetry
of the Universe thanks to low-energy {\it CP} violation
alone~\cite{davidson2,us2,many}.
This intriguing scenario is not viable in the unflavored picture
and makes the search for {\it CP} violation in the neutrino
sector even more exciting.

On the other hand, accounting for flavor effects
does not help to solve the problem of the large reheating temperatures
($T\gtrsim 3\times 10^{9}\,{\rm GeV}$) required
to have successful leptogenesis when a hierarchical
RH neutrino spectrum is assumed \cite{us2}. This problem
can only be solved, within the minimal scenario,
going beyond the hierarchical limit for the spectrum
of RH neutrino masses \cite{pilaftsis,beyond}.

Another important implication of flavor effects concerns the
role  played by the absolute neutrino mass scale in determining the
asymmetry. In the unflavored case there is a stringent upper bound
on the lightest active neutrino mass, $m_1 \leq 0.1$ eV~\cite{neutrinobound}.
It has been claimed in \cite{davidson1} that this bound disappears when flavor
effects are accounted for using the `fully flavored regime',
where a set of classical Boltzmann equations can be still used.
However, the right condition for the fully flavored regime to be valid
has been found to be more restrictive \cite{us3}.
This implies that a conclusive answer to this issue
and a precise value of the bound
requires additional investigation and in particular a full
quantum kinetic description of the lepton and anti-lepton
density matrix evolution.

\section{Fully flavored regime}

Leptogenesis relies on the
see-saw mechanism, the simplest way to explain neutrino masses.
The Standard Model Lagrangian is augmented with a neutrino Yukawa-type
term ($h \bar{L}N\F$) and a Majorana mass term for the
heavy right-handed (RH) neutrinos, whose masses are denoted
with $M_1 \leq M_2 \leq M_3$.
Let us assume that a `fully flavored regime' holds.
This means that the leptons are fully projected onto the flavor
base and are described in terms of flavor eigenstates.
We will discuss later on the condition for the
validity of the fully flavored regime.
Since we will consider RH neutrino masses larger than
$\sim 10^9$ GeV, the muon Yukawa interactions are not
effective and we can limit our discussion to the two-flavor case
($\t$ and a combination of $\m$ and $e$, which we shall denote `$e\m$').
The individual flavored asymmetries, $\D_{\a}\equiv B/2-L_{\a}$,
have to be tracked separately and the final asymmetry has to be
calculated as a sum $N_{B-L}^{\rm f}=\sum_{\a}\,N_{\D_{\a}}$,
where with $N_X$ we indicate the $X$-abundance per number of RH
neutrinos in ultra-relativistic equilibrium.
The set of flavored equations can then be written as

\bea
{dN_{N_1}\over dz} & = & -D_1\,(N_{N_1}-N_{N_1}^{\rm eq}),\label{eq1} \\
\nonumber
{dN_{\D_{\a}}\over dz} & = &
\ve_{1\a}\,D_1\,(N_{N_1}-N_{N_1}^{\rm eq})
-P_{1\a}^{0}\,W_1^{\rm ID}\,N_{\D_{\a}}\label{eq2} \, ,
\eea
where $z\equiv M_1/T$ and, for simplicity,
we are taking into account only decays and inverse decays.
The decay term is given by $D_1\equiv \G_{\rm D 1}/(H z) \propto K_1$,
where $\G_{\rm D 1}$ is the total decay rate of $N_1 \to L \F^{\dagger}(\bar{L} \F)$,
$H$ is the expansion rate and $K_1$ is the decay parameter;
the inverse decay wash-out rate is given by
$W_1^{\rm ID}\equiv \G_{\rm ID}/(2 H z)\propto K_1$ where
$\G_{\rm ID}=\G_{\rm D 1} N_{N_1}^{\rm eq}$ is the inverse decay rate.

The projectors $P_{1\a}$ give the probability
 that the lepton state coming from the decay of the heavy neutrino
 has a flavor $\a$, such that they add up to 1 and
 the upper script `0' indicates tree level.
 There are two main differences compared to the usual unflavored case.
 1) The individual {\it CP} asymmetries are not simply proportional
    to the total $C\!P$ asymmetry $\ve_1$ but they contain an additional
    contribution such that
 $\ve_{1\a}= P_{1\a}^{0}\ve_{1} + \D P_{1\a}/ 2$ \cite{bcst,nardi},
 where the differences $\D P_{1\a}\equiv P_{1\a} - \bar{P}_{1\a}$ arise
 from loops and they are naturally of order $\mathcal{O}(\ve_1)$.
2) The $\a$-flavor wash-out term is reduced
by the projector $P_{1\a}^{0}$.

\section{Implications of flavor effects}

Let us now discuss how the leptogenesis predictions
get modified in the fully flavored regime.
\begin{enumerate}
\item In the unflavored case, assuming a hierarchical spectrum
of RH neutrinos, $M_2\gtrsim 3\,M_1$,
successful leptogenesis implies a stringent lower bound on
$M_1$ and on the reheating temperature ~\cite{ibarra}. This is given by
$M_1\,(T_{\rm reh})\gtrsim 3\,(1.5)\times 10^{9}\,{\rm GeV}$
at the onset of the strong wash-out regime
for $K_1\gtrsim 3$, where there is no dependence on the initial
conditions and by $M_1,T_{\rm reh}\gtrsim 4\times 10^{8}\,{\rm GeV}$
for an initial thermal abundance and for $K_1\ll 1$.
This lower bound  conflicts with the upper bound coming from the
avoidance of the gravitino problem within a super-symmetric version.
Unfortunately, flavor effects do not help to solve this problem \cite{us2}.
 It is however true
that at fixed $K_1 \gg 1$, there can be a relaxation
that is only a factor $\sim 2$ in the `democratic case', when
$P_{\t}\simeq P_{e\m}\simeq 1/2$, while it can be much larger in a
$\a$-dominated scenario, especially when the
low-energy phases are switched on \cite{us2}.

\item Flavor effects disclose a new intriguing possibility,
a scenario of leptogenesis where the asymmetry production
stems uniquely from non-zero low-energy
{\it CP}-violating phases~\cite{davidson2,us2,many}.
 This is possible because the additional contribution to the
 individual {\it CP} flavored asymmetries, the $\D P$'s,
 depends on the PMNS mixing matrix. On the contrary, it is well
 known that the dependence
 in the total $C\!P$ asymmetry cancels out
 by unitarity. Therefore, even though the total
 {\it CP} asymmetry $\ve_{1}=\sum_{\a}\ve_{1\a}$ 
  vanishes, the individual {\it CP} asymmetries in general do not.
However, in this scenario and for a hierarchical
heavy neutrino spectrum, successful leptogenesis is obtained mostly
in the weak wash-out regime where there is dependence on the
initial conditions \cite{us2}. This problem can be circumvented
going beyond the assumption of a hierarchical heavy neutrino spectrum
\cite{preparation}.

\item Besides the extreme case of leptogenesis from low-energy
 phases, flavor effects can yield large deviations
from the unflavored case also in other situations,
in particular in the so-called `one-flavor dominated scenario' ~\cite{us2}.
This is realized when one projector is much smaller than unity
but nevertheless, thanks to the $\D P$ contribution,
the two $C\!P$ asymmetries are comparable with each other.
This scenario is not as exceptional as one could think \emph{a priori},
especially when going toward a degenerate light neutrino
spectrum and/or when low-energy phases are turned on.

\item There exists an upper bound on the individual {\it CP}
asymmetries, proportional
to the absolute neutrino mass scale~\cite{davidson1}
\be
|\ve_{1\a}| <
\bar{\ve}(M_1)\sqrt{P^0_{1\a}}{m_3 \over m_{\rm atm}} {\rm max}_j[|U_{\a j}|],
\ee
where $\bar{\ve}(M_1)\equiv 3 M_1 m_{\rm atm}/(16 \p v^2)$. This is
in contrast with the upper bound on the total $C\!P$ asymmetry
that is inversely proportional. The absolute value of the
individual {\it CP} asymmetries can thus become much larger
than the one of the total $C\!P$ asymmetry
when a quasi-degenerate neutrino spectrum is considered.
This leads to the conclusion that
the stringent upper bound on the neutrino masses holding in the
unflavored regime, $m_1 \leq 0.1$ eV~\cite{neutrinobound},
disappears in the fully flavored regime ~\cite{davidson1}.

\item
The possibility to circumvent the neutrino
mass upper bound motivates a careful analysis of the
condition of validity of the fully flavored regime.
A necessary condition is that the tauon Yukawa interactions
must be in equilibrium, i.e. $\G_{\t}\gtrsim H$,
and this translates into $T\lesssim 10^{12}$~GeV~\cite{nardi,davidson1}.
However, this condition is actually
not sufficient when $K_1\gg 1$~\cite{us3}. More restrictively, one has
to impose that the charged Yukawa interactions have to be
faster than the RH neutrino inverse decays. This condition is
maximally restrictive in the case of one-flavor dominance, affecting the
possibility to evade the neutrino masses upper bound.
The reason is simply that when $m_1$ increases in the
quasi-degenerate spectrum, $K_1$ -- and therefore
the inverse decay rate -- increases as well,
making the condition for the validity of the fully flavored regime
more restrictive. This results into an inability
of the fully flavored regime to solve the issue
whether the upper bound holding in the unflavored
case is circumvented or not and into the necessity
of a full quantum kinetic description
of the lepton density matrix evolution. Notice that
an upper bound $m_1 \lesssim 2\,{\rm eV}$ holding in the fully flavored regime
and consistent with the more restrictive condition that 
we have just discussed has been found \cite{us3,desimone},
but it is not saturated and moreover affected by large
uncertainties and therefore not conclusive.
\end{enumerate}

\section{Conclusions}
Flavor effects modify the conventional picture
in a very interesting way. The most striking
consequence is that a scenario where only the low energy
phases are responsible for $C\!P$ violation becomes viable.
On the other hand,
flavor effects do not solve the problem
of large reheating temperatures when a hierarchical RH neutrino
spectrum is assumed. It is still not obvious
if the neutrino mass upper bound is evaded.
A full quantum kinetic description is needed in this respect.

\textbf{Acknowledgments}
We would like to thank the organizers of NOW 2006 for a very pleasant and
successful workshop, and S.~Petcov for interesting discussions on the subject.

\end{document}